\documentclass[prd,twocolumn,showpacs,preprintnumbers,amsmath,amssymb,floatfix,nofootinbib]{revtex4}

\usepackage{subfigure,graphicx,epsfig,amsmath,amsfonts,amssymb,xcolor,slashed,ulem}

\newcommand{\be}{\begin{equation}}
\newcommand{\ee}{\end{equation}}
\newcommand{\ba}{\begin{eqnarray}}
\newcommand{\ea}{\end{eqnarray}}
\newcommand{\nn}{\nonumber}
\newcommand{\f}{f_1(1285)}
\newcommand{\mev}{\textrm{ MeV}}
\newcommand{\gev}{\textrm{ GeV}}

\begin{document}

\title{ Triangle singularity in $\tau \to  f_1(1285)\pi\nu_\tau $ decay}

\author{E.~Oset}
\affiliation{Departamento de
F\'{\i}sica Te\'orica and IFIC, Centro Mixto Universidad de
Valencia-CSIC Institutos de Investigaci\'on de Paterna, Aptdo.
22085, 46071 Valencia, Spain}

\author{L. Roca}
\affiliation{Departamento de F\'isica, Universidad de Murcia, E-30100 Murcia, Spain}

\date{\today}

\begin{abstract}

We show that the $\tau^-$ decay into $f_1(1285)\pi^-\nu_\tau $ is dominated by a triangle loop mechanism with $K^*$, $\bar K^*$ and $K$ (or $\bar K$) as internal lines, which manifests a strong enhancement reminiscent of a nearby singularity present in the narrow $K^*$ limit. The $f_1(1285)$ is then produced by its coupling to the $K^*\bar K$ and $\bar K^*K$ which is obtained from a previous model where this resonance was dynamically generated as a molecular $K^*\bar K$ (or $\bar K^*K$) state using the techniques of the chiral unitary approach. We make predictions for the $f_1 \pi$ mass distribution which significantly deviates from the phase-space shape, due to the distortion caused by the triangle singularity. We find a good agreement with the experimental value within uncertainties for the integrated partial decay width, which is a clear indication of the importance of  the triangle singularity in this decay and supports the dynamical origin of the $\f$ as a $K^*\bar K$ and $\bar K^* K$ molecular state.

\end{abstract}

\maketitle

\section{Introduction} \label{sec:intro} 

The hadronic decays of the $\tau$ lepton have traditionally been advocated as one of the best frameworks to study the low energy sector of the strong interaction and the hadronic weak currents. (See Ref.~\cite{Lafferty:2015hja} for a quick review).
The $\tau$ is the only lepton heavy enough to decay into hadrons, and only mesons can be produced since otherwise it would require at least a baryon-antibaryon which have not enough phase space available. Furthermore there is only energy to produce up to strange mesons and thus $\tau$ decays represent a clean frame to study light flavor meson dynamics.
Inclusive reactions allow for precise determination of parameters of the standard model and exclusive ones for the study of the mesonic spectrum and resonance parameters \cite{Pich:2013lsa}. For instance, the $\tau\to\pi\pi\pi\nu$ decay is a priceless reaction to obtain the axial-vector spectral function \cite{Schael:2005am,Davier:2013sfa}.


In this work we are concerned with the little studied decay channel, $\tau \to  f_1(1285)\pi\nu_\tau $. This decay is specially important because it explores the moderate energy region of $f_1\pi$ invariant mass between $m_{f_1}+m_{\pi}\simeq 1420\mev$ and $m_\tau=1777\mev$ which is too high for standard chiral perturbation theory but too low for perturbative QCD.
Usually theoretical studies have resorted to phenomenological models, like \cite{Li:1996md}, where ChPT is supplemented with the inclusion of vectors and axial-vectors through VMD, or like Nambu-Jona-Lasinio (NJL) models \cite{Calderon:2012zi,Vishneva:2014lla,Volkov:2018fyv}. 
In the previous works it was concluded the importance of the role played by the  $a_1(1260)$ and the $a_1(1640)$ axial-vector resonances.

In the present work we approach the $f_1(1285)\pi^-\nu_\tau $ decay from a different point of view and we justify that it is dominated by a  mechanism which in principle should be small but which is enhanced by a nearby singularity of the loop involved. This mechanism  (see Fig.~\ref{fig:triang}) is the triangle loop formed by a $K^*$ and a $\bar K^*$ produced from the hadronic weak current and then the $K^*$ (or the $\bar K^*$) coupling to a  $K$ (or $\bar K$) and  a $\pi$ and the $ \bar K^*$ (or the $K^*$) merges with the previous kaon to produce the $\f$.
The study of the singularities produced by triangle diagrams was brought up by Landau \cite{landau} but has gained renewed interest in view of the increased hadronic experimental information
\cite{Liu:2015taa,Guo:2017jvc}. The triangle diagrams, of the kind of the one shown in Fig.~\ref{fig:triang}, produce a singularity only if the processes involved in the loop  vertices can occur at the classical level, (Coleman-Norton theorem \cite{ncol}). If the particles inside  the loop have  a finite width, the singularity takes the shape of a broad peak and it can be misinterpreted as an actual resonance. It can be an important reason for resonance misidentifications together with cusp enhancements \cite{Guo:2017wzr}.
However, till very recently not many experimental examples where the triangle singularity played a crucial role had been shown up. In \cite{Liu:2015taa,mikha,aceti} it was shown
that a peak experimentally observed at COMPASS \cite{compass}
is essentially produced due to a triangle singularity involving the decay of the $a_1(1260)$ resonance.
In other cases an a priori suppressed mechanism, like the 
isospin violating $\eta(1405) \rightarrow \pi\,f_0(980)$, is enhanced thanks to a triangle singularity \cite{wu2,wu1,wu3}.
An example of triangles producing peaks not associated at all to a resonance is the peak obtained in the  $\gamma p\rightarrow K^+\Lambda(1405)$ reaction \cite{wang} or in  $\gamma p \to \pi^0 \eta p$ \cite{shuntaro}.
Even in the heavy quark sector triangle singularities has been predicted to produce peaks or give an abnormally large contribution
\cite{Sakai:2017hpg,Pavao:2017kcr,Sakai:2017iqs}.
Further recent examples where the triangle singularity has played a crucial role for specific processes can be found in 
 Refs.~\cite{hanhart,achasov,Lorenz,adam1,adam2,Roca:2017bvy,daris}.

In the present work we show that the $\tau \to  f_1(1285)\pi\nu_\tau $
is also dominated by this kind of enhancement from a triangle diagram. Actually, considering only this mechanism, we obtain a good agreement with the experimental branching ratio. Furthermore, the strength of the process is crucially determined by the coupling of the $\f$ to $K^*\bar K$ and $\bar K^*K$ which we obtain from a previous model \cite{Roca:2005nm} where the $\f$ was dynamically generated using the techniques of the chiral unitary approach. Thus the value of the coupling is a genuine prediction of that picture and the fact that the branching ratio agrees with the experimental value within uncertainties supports the dynamical (or molecular) picture of the $\f$.


\section{Formalism}

\begin{figure}[tbp]
     \centering{
          \includegraphics[width=.95\linewidth]{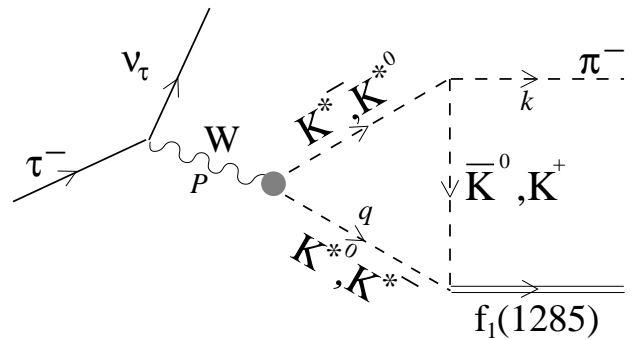}}\\
    \caption{Triangular mechanism for the $\tau \to  f_1(1285)\pi^-\nu_\tau $ decay}\label{fig:triang}
\end{figure}

We first justify why we expect the $\tau^- \to  f_1(1285)\pi^-\nu_\tau $ decay to be dominated by the triangle diagram shown in Fig.~\ref{fig:triang}.
First of all, the $\f$ resonance must be produced from $K^*\bar K$ and $\bar K^*K$ since this resonance was dynamically generated from this channels in ref.~\cite{Roca:2005nm}. In general, the work in ref.~\cite{Roca:2005nm} evaluated the S-wave interaction of the octet of pseudoscalar mesons ($P$) and the nonet of vector mesons ($V$) with the only input of a lowest order chiral Lagrangian for the $VP\to VP$ interaction. The implementation of unitarity in coupled channels, allowed to obtain dynamically most of the lowest axial-vector resonances which are identified by poles in unphysical Riemann sheets of the unitarized scattering amplitudes.
Implementation of higher order terms in the kernel (potential) in this chiral unitary approach was found to have negligible effects \cite{Zhou:2014ila}.
From the residues at the poles, the couplings to the different channels can be obtained. In particular, and of interest for the present work, in the isoscalar channel a pole was found at $\sqrt{s}=1288\mev$ in the real axis in the physical sheet from the scattering of the  positive G-parity $\frac{1}{\sqrt{2}}(\bar K^*K - K^*\bar K)$ state\footnote{Note the different $G-$ and $C-$parity prescription with respect to ref.~\cite{Roca:2005nm}. Here we use $C|P\rangle=|P\rangle$ and $C|V\rangle=-|V\rangle$ and $|K^-\rangle$ and $|{K^*}^-\rangle$ isospin states are $-|I=\frac{1}{2},I_3=-\frac{1}{2}\rangle$  and thus the $\f$ wave function is $|f_1\rangle=-\frac{1}{2}|K^+{K^*}^-
+K^0 {{\overline K}^*}^0
-K^-{K^*}^+ -\overline K^0 {K^*}^0\rangle$.},   which is the only allowed positive G-parity combination of a vector and a pseudoscalar meson with 0 isospin and 0 strangeness. The position of the pole is 100~MeV below the $K^ *\bar K$ threshold and thus it corresponds to a bound state. The coupling obtained in \cite{Roca:2005nm} for the $\f$ to  
$\frac{1}{\sqrt{2}}(\bar K^*K - K^*\bar K)$ was $g_{f_1}=7230\mev$, if the $VP$ loop functions are regularized using dimensional regularization. 
However, for the reasons explained below, we need the equivalent three-momentum  cutoff if using a cutoff regularization scheme. This three-momentum cutoff, $\Lambda$, turns out to be $\Lambda=1000\mev$ in order to get the $f_1$ pole at the experimental mass value. With the cutoff regularization we obtain now that the $g_{f_1}$ coupling is $g_{f_1}=7475\mev$. Altogether we can assign for the value of that coupling $g_{f_1}=7350\pm 130\mev$.
Therefore, in this model the $\f$ resonance can be interpreted as a molecular $\bar K^* K$ and $K^*\bar K$ state and thus in the $\tau \to  f_1(1285)\pi^-\nu_\tau $ decay it must be dominantly produced from a $\bar K^* K$ and $K^*\bar K$ through the diagram depicted in Fig.~\ref{fig:triang}. Other diagrams with different combinations of $K^ *$'s and $K$'s inside the triangle require $VVP$ or $PPP$ anomalous vertices or require the coupling of the $W^-$ to $\bar K K$ which is an order of magnitude smaller than the coupling to $\bar K^* K^*$, as can be deduced from the ratio of the phase space allowed in these decays and the experimental decay value. Furthermore, all those other possible diagrams are not enhanced by the triangle singularity explained below.

The triangle diagrams of the kind of that shown in Fig.~\ref{fig:triang} are usually small except for particular kinematic conditions which are fulfilled in the present case. As shown in \cite{Bayar:2016ftu,ncol}, the triangle presents a singularity when the particles in the loop are collinear and the process can occur at the classical level (Coleman-Norton theorem).
In other words, the singularity happens if the three particles in the loop can go on-shell and the momentum of the lowest $K^*$ and the $K$ inside the loop are parallel and the $K$ meson moves faster than the lowest $K^*$ so that it can catch up with the $K^*$ to rescatter.
Mathematically, this occurs
 for a value of the incoming energy $\sqrt{s}$ (the $\f\pi$ invariant mass) which is the solution of the equation \cite{Bayar:2016ftu}
\be
q_{on}-q_{a^-}=0,
\label{eq:qomenosqa}
\ee
where $q_{on}$ is the on-shell momentum of the $K^*$ in the $W^-$ (or $\tau\nu_\tau$) rest frame with total energy $\sqrt{s}$, and 
$q_{a^-}$ 
corresponds to the solution for $\f\to K^* \bar K$ in the $W^-$ rest frame which has negative imaginary part $(-i\varepsilon)$, with $\vec q$ and $\vec k$ antiparallel.
 (See reference \cite{Bayar:2016ftu} to see explicit expressions for $q_{on}$ and $q_{a^-}$ and a justification for Eq.~\eqref{eq:qomenosqa} as well as for further discussion on the origin of the singularity).
 It is worth noting that the $\f$ has a mass about 100$\mev$ below the $K^*\bar K$ threshold and then the singularity cannot explicitly occur since the solution of Eq.~\eqref{eq:qomenosqa} requires the invariant mass of the  $\f$  to be larger than the threshold of the lowest $K^*$ and the $K$. However, for a value of the invariant mass of the $\f$ leg just a few MeV above the $K^* \bar K$ threshold, a singularity occurs for  a value for $\f\pi$ invariant mass ($M_{f_1\pi}$)  slightly above 1800~MeV. That value of $M_{f_1 \pi}$ is never reached in the present $\tau$ decay, since the maximum value of  $M_{f_1\pi}$ is $m_\tau=1777\mev$, but the pole is close enough to the allowed phase-space as to influence the  $f_1\pi$ mass distribution and benefit from the proximity to the singularity.
In any case, the previous semiqualitative discussion about the importance of the triangle mechanism of Fig.~\ref{fig:triang} needs to be backed up with its proper quantitative evaluation which we carry out in the following.

\begin{figure}[tbp]
     \centering{
          \includegraphics[width=.95\linewidth]{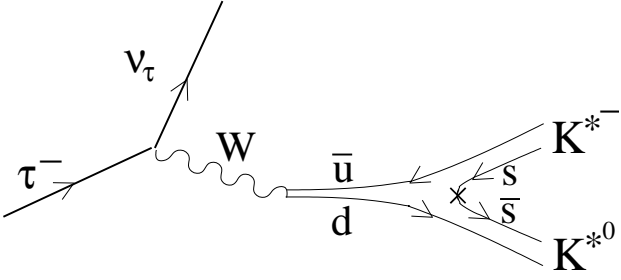}}\\
    \caption{$K^* \bar K^*$ production at the quark level}\label{fig:diagqq}
\end{figure}
 
Let us address, next, the evaluation of the  $\tau\to\nu_\tau K^* \bar K^*$ part. 
The  $K^* \bar K^*$ production is assumed to proceed first from the Cabibbo dominating $\bar u d$ production from the $W^-$ which then hadronizes producing an $s \bar s$ with quantum numbers, $^3P_0$, so that the pair has the vacuum quantum numbers, see Fig.~\ref{fig:diagqq}.
The production is dominant for final  $K^* \bar K^*$ in S-wave and $J=0$. This implies that the initial $\bar u d$ pair must be in $L=1$, $S=1$ to match the parity and angular momentum of the final $K^* \bar K^*$ pair.

The $\bar u d$ production vertex is proportional to $
\langle \bar u_d|\gamma^\mu(1-\gamma_5)|v_u\rangle
$. Therefore, up to global constants, the hadronic part accounting for the $W^-\to K^* \bar K^*$ must be
\be
H^\mu=\langle \bar u_d|\gamma^\mu(1-\gamma_5)|v_u\rangle.
\label{eq:qqmat}
\ee
In the frame of reference  where the $W^-$ is at rest, both $\bar u$ and $d$ quarks have a small momentum since they will have an invariant mass close to $2m_{K^*}$. Therefore, in the limit of small  momenta of the quarks and for the total spin $S=1$ combination, the quark matrix element of Eq.~\eqref{eq:qqmat} reduces to (see Appendix \ref{appendix:app} for  explicit details of the evaluation)
\be
H^\mu=\sum_{s_1} \langle \frac{1}{2} s_1 | \sigma^i | \frac{1}{2} -s_2\rangle \,{\cal C}\left(\frac{1}{2}, \frac{1}{2}, \ 1 ; s_1, \ -s_2\right) (-1)^{\frac{1}{2}+s_2}.
\label{eq:qqmatred}
\ee
with $s_1$ ($s_2$) the spin third component of $d$ ($\bar u$) and ${\cal C}(...)$ a Clebsch-Gordan coefficient. Note that $H^\mu$ has only space components in the present case. The phase in Eq.~\eqref{eq:qqmatred} comes from particle-hole conjugation to account for the antiquark state.

For the leptonic current, $L^\mu$, accounting for the part at  the  left of the
$W^-$ boson, we have, up to constants,
\be
L^\mu=\langle \bar u_\nu(p')\gamma^\mu(1-\gamma_5)| u_\tau(p)\rangle,
\ee
with $p$ ($p'$) the $\tau$ ($\nu_\tau$) momentum and which satisfy that $\vec p=\vec p\,'$ in the $W^-$ rest frame.

When evaluating the $\tau$ decay width, we will need the sum over the lepton polarizations which will give the standard result
\be
\overline \sum L^\mu {L^\nu}^\dagger= 4\left( p^\mu p'^\nu
 +p'^\mu p^\nu - g^{\mu\nu}p'\cdot p+i \epsilon^{\alpha\mu\beta\nu}p'_\alpha p_\beta\right).
 \label{eq:LL}
\ee
For the evaluation of the decay width we will need the contraction with the hadronic part $H^\mu$ from Eq.~\eqref{eq:qqmatred}, which after some algebra (see Appendix \ref{appendix:app}) gives
\be
\overline \sum L^\mu {L^\nu}^\dagger H_\mu H_\nu^\dagger=
24\left( E_\tau E_\nu -\frac{\vec p\,^2}{3}\right),
\label{eq:contr}
\ee
where $E_\tau$ and   $E_\nu$, are the $\tau$ and the $\nu_\tau$ energies in the $W^-$ boson rest frame.

Therefore, for the amplitude $\tau\to\nu_\tau {K^*}^0 {K^*}^-$, we can effectively use
\be
t_{\tau\to\nu_\tau  {K^*}^0 {K^*}^-}=
C\left(E_\tau E_\nu -\frac{\vec p\,^2}{3}\right)^{\frac{1}{2}}
\vec\epsilon(K^*)\cdot\vec\epsilon\,'(\bar K^*).
\label{eq:tkstar}
\ee
with $\vec\epsilon(K^*)$ and $\vec\epsilon\,'(\bar K^*)$ the polarization vectors of the $K^*$ and $\bar K^*$.
In Eq.~\eqref{eq:tkstar}, $C$ is a constant which can be obtained from experimental data of the partial decay width of 
 $\tau\to\nu_\tau K^* \bar K^*$.
Indeed, using the amplitude of Eq.~\eqref{eq:tkstar}, the invariant mass distribution of the partial decay width  for
  $\tau\to\nu_\tau K^* \bar K^*$ is
  \be
  \frac{d\Gamma_{\tau\to\nu_\tau K^* \bar K^*}}{dM_{K^* \bar K^*}}=\frac{1}{(2\pi)^3}\frac{\bar p_\nu \,\widetilde p_{K^*}}{4 m_\tau^2}
3\,C^2\left( E_\tau E_\nu -\frac{\vec p\,^2}{3}\right)  
\label{eq:GKsKS}
\ee
where $\bar p_\nu$ is the $\nu_\tau$ three-momentum in the $\tau$ rest frame, 
$\bar p_\nu=\lambda^{1/2}(m_\tau^2,0,M_{K^* \bar K^*}^2)/(2 m_\tau)$
 and $\widetilde p_{K^*}$ the $K^*$ three-momentum in the $K^* \bar K^*$ rest frame,  $\widetilde p_{K^*}=\lambda^{1/2}(M_{K^* \bar K^*}^2,M_{K^*}^2,M_{K^*}^2)/(2M_{K^* \bar K^*})$.
 Note, however, that there would be no available phase space for the decay  
$\tau\to\nu_\tau K^* \bar K^*$ if the $K^*$ was infinitely narrow. Thus the $\tau\to\nu_\tau K^* \bar K^*$ takes place only because of the tails of the $K^*$ and $\bar K^*$ distributions. 
Therefore, in order to evaluate the partial width $\Gamma_{\tau\to\nu_\tau K^* \bar K^*}$ we need to account for the finite $K^*$ widths. We do this by folding the width coming from Eq.~\eqref{eq:GKsKS}  with the spectral distribution of the $K^*$ and $\bar K^*$:

\begin{align}
\Gamma_{\tau\to\nu_\tau K^* \bar K^*}=& \int_{(m_\pi+m_K)^2}^{(m_\tau-m_\pi-m_K)^2} dm_1^2\int_{(m_\pi+m_K)^2}^{(m_\tau-m_1)^2} dm_2^2 \nn \\
&\times\frac{-1}{\pi}Im D_{K^*}(m_1)\frac{-1}{\pi}Im D_{K^*}(m_2)
\nn \\
&\times\int_{m_1+m_2}^{m_\tau}dm_{12}
\frac{d\Gamma_{\tau\to\nu_\tau K^* \bar K^*}(m_1,m_2)}{d m_{12}}
\label{eq:convol}
\end{align}
where $m_1$ and $m_2$ have to be used as the masses of the $K^*$ and $\bar K^*$ in Eq.~\eqref{eq:GKsKS}. Note that Eq.~\eqref{eq:convol}  essentially accounts for the five-body phase space for the decay  
$\tau\to\nu_\tau K^* \bar K^*\to \nu_\tau K \pi \bar K \pi$.
In Eq.~\eqref{eq:convol}, $D_{K^*}(m)$ stands for the $K^*$ propagator. Since the $\tau\to\nu_\tau K^* \bar K^*$ only sees the lowest part of the $K^*$ tails, it is very important to have a good parametrization of the $K^*$ spectral distribution from the $K \pi$ threshold on. Therefore we use an energy dependent width in the propagators considering also the Blatt-Weisskopf penetration factors \cite{blattweisskopf} to take into account the form factor of the $K^*$ decay into $K \pi$:
\be
D_{K^*}(m)=\frac{B'_1(m)}{m^ 2-m_{K^*}^2+i m_{K^*} \Gamma(m)}
\ee
with
\be
\Gamma(m)=\Gamma_o \frac{m_{K^*}}{m}\frac{p^3(m)}{p^3(K^*)} B'_1(m)^2
\ee
where $\Gamma_o$ is the total width of the $K^*$, $p(m)$ the $K$ or $\pi$ momentum in the $K^*$ rest frame for a $K^*$ invariant mass $m$,   and  $B'_1(m)$ is the p-wave Blatt-Weisskopf barrier penetration  factor \cite{blattweisskopf} given by
\be
B'_1(m)=\left(\frac{1+(R\, p(m_{K^*}))^2}{1+(R\, p(m))^2}\right)^{1/2}
\label{eq:blatt}
\ee
In Eq.~\eqref{eq:blatt}, $R$ stands for the range parameter of the $K^*$ for which we use an average of the values reported by the PDG \cite{pdg}, $R=3.2\pm 1.0~\textrm{GeV}^{-1}$.

The experimental  data for $\tau\to\nu_\tau K^* \bar K^*$ to get the global normalization of the $K^*\bar K^*$ vertex, ${C}$, can be obtained from the branching ratios quoted in the PDG \cite{pdg}, $\textrm{BR}(\tau^-\to\nu_\tau {K^*}^-K^0\pi^0\to\nu_\tau\pi^-K^0_SK^0_S\pi^0)=(1.08\pm0.14\pm0.15)\times 10^{-5}$ and $\textrm{BR}(\tau^-\to\nu_\tau K^-K^+\pi^ -\pi^0)=(6.1\pm 1.8)\times 10^{-5}$, which provide a value $\textrm{BR}(\tau^-\to\nu_\tau {K^*}^-{K^*}^0)=(2.1\pm 0.5)\times 10^{-4}$.

Once we have the $\tau\to\nu_\tau K^* \bar K^*$ part,
we next proceed to evaluate the rest of the diagram in Fig.~\ref{fig:triang}.
In addition to the $K^*\bar K^*$ vertex and the $\f K^*\bar K$ explained above, we need also the $K^* K\pi$ vertex that we obtain from the  $VPP$  Lagrangian
\begin{equation}
\mathcal{L}_{VPP}=-ig\langle V^{\mu}[P,\partial_{\mu}P]\rangle\ ,
\label{eq:VPPlag}
\end{equation}
where $P$ ($V$) are the usual $SU(3)$ matrices containing the pseudoscalar (vector) mesons. The coupling $g$ is of the order of $m_\rho/(2 f)=4.14$, with $f=93\mev$ the pion decay constant, but we can fine tune the value for $K^* K \pi$ vertex from the experimental value of the decay width of $K^*\to K \pi$, and we get $g=4.31\pm 0.10$.

The sum of amplitudes for the diagrams of 
Fig.~\ref{fig:triang},
considering the $VPP$ couplings of  Fig.~\ref{fig:triang} and the $K\bar K^*$ components of the $f_1$ wave function,
 takes the form, in the $W^-$ (or $f_1\pi$) rest frame ($\vec P=0$),

\begin{align}
t_{\tau^- \to  f_1(1285)\pi^-\nu_\tau }=-{C}\,g\, g_{f_1}\left( E_\tau E_\nu -\frac{\vec p\,^2}{3}\right) \vec \epsilon(f_1)\cdot\vec k \,t_T,
\label{eq:tall1}
\end{align}
from where the final expression for the mass distribution of the partial decay width is:
\be
\frac{d\Gamma_{\tau\to f_1\pi \nu_\tau}}{dM_{f_1\pi}}=\frac{1}{(2\pi)^3}\frac{\bar p_\nu \,\widetilde p_{\pi}}{4 m_\tau^2}
\,{C}^2\left( E_\tau E_\nu -\frac{\vec p\,^2}{3}\right) 
\vec k^2 |t_T|^2
\label{eq:BRall1}
\ee
In Eqs.~\eqref{eq:tall1} and \eqref{eq:BRall1}, $t_T$ stand for the
triangle loop function which in the present case reads
\begin{align}
t_T=&i\int \frac{d^4q}{(2\pi)^4}
\, 
\frac{2+ \vec k\cdot\vec q/\vec k\,^2}{(P-q)^2-m_1^2+i\varepsilon}\,\nn\\
&\times\frac{1}{q^2-m_2^2+i\varepsilon}\,
\frac{1}{(P-q-k)^2-m_3^2+i\varepsilon},
\label{eq:tTd4}
\end{align}
where the labels 1, 2, 3 stand for the particles on the upper side, lower side and right side respectively of the triangle in Fig.~\ref{fig:triang}.
After performing the  integration in $q^0$, the amplitude in Eq.~(\ref{eq:tTd4})
takes the form
\begin{align}
t_T=&\int\frac{d^3q}{(2\pi)^3}\frac{2+ \vec k\cdot\vec q/\vec k\,^2}{8 \omega_1 \omega_2 \omega_3}\,
\frac{1}{k^0-\omega_1-\omega_3}\,
\frac{1}{P^0-k^0+\omega_2+\omega_3}\,\nn\\
&\times\frac{1}{P^0-k^0-\omega_2-\omega_3+i\varepsilon}\,
\frac{1}{P^0-\omega_1-\omega_2+i\varepsilon}\nn\\
&\times\left[2P^0 \omega_2+2k^0 \omega_3-2(\omega_2+\omega_3)(\omega_1 +\omega_2 +\omega_3)\right],
\label{eq:tTd3}
\end{align}
where
$\omega_{1,2}=\sqrt{m_{1,2}^2+\vec q\,^2}$, 
$\omega_3=\sqrt{m_3^2+(\vec k+\vec q)^2}$, $P^0=M_{f_1\pi}$ and $k=\lambda^{1/2}({P^0}^2,m_\pi^ 2,m_{f_1}^2)/(2 P^0)$ is the $\pi^-$ momentum in the $\pi f_1$ rest frame.
The integral in Eq.~\eqref{eq:tTd3} is convergent, however we have to include a three-momentum cutoff which is the same as the one needed to regularize the $K^*\bar K$ loop in the dynamical generation of the $f_1(1285)$ within the chiral unitary approach of 
Ref.~\cite{Roca:2005nm}. Indeed, it was justified in Ref.~\cite{Gamermann:2009uq} that if a three-momentum cutoff is used in the $VP$ potential, it translates into the cutoff needed to regularize the $VP$ loop function in the evaluation of the unitarized amplitude. Furthermore it is also shown in Ref.~\cite{Gamermann:2009uq} that the final $VP\to VP$ full amplitude is affected by the same cutoff for the external momenta in the resonance rest frame. Since the $g_{f_1}$ coupling we use is obtained within that model, then we have
to limit the momentum allowed to enter the $f_1$ with the same cutoff, which is $\Lambda\equiv 1000\mev$ in the present case \cite{Roca:2005nm}. Since $\Lambda$ refers to the three-momentum in the $\f$ rest frame, we have to boost the $q$ momentum to that frame, which gives
$
{q^*}=({{q^*}^0}^2-m_{K^*}^2)^{1/2},
$
with
$
{q^*}^0=[(P^0-k^0)q^0+ \vec q\cdot \vec k]/m_{f_1}.
$
Therefore we have to add to the integrand of Eq.~\eqref{eq:tTd3} the factor $\Theta(\Lambda-q^*)$, where $\Theta$ is the step function.

 As mentioned in the discussion around Eq.~\eqref{eq:qomenosqa}, the function $t_T$ has a singularity if  Eq.~\eqref{eq:qomenosqa} has a solution. For the value of the masses of the present triangle there would be a singularity, located a bit above 1800$\mev$, if the mass of the  $\f$ was about 1390$\mev$, which is not the actual value of the $\f$ but is not to far away and hence the amplitude can still have an important enhancement even though there is not an exact singularity.
Furthermore, the singularity only happens strictly in the limit of narrow resonances in the loop. The usual effect of considering the finite widths of the resonances inside the triangle results in a broadening of the peak and thus the singularity turns into a bump.
In order to take into account the finite widths of the $K^*$ mesons inside the loop, we can add a factor $i\Gamma_{K^*}/2$ in the denominators of Eq.~(\ref{eq:tTd3}) for each $K^*$ that can be on-shell in the integration of the loop.
 \begin{figure}[tbp]
     \centering{
          \includegraphics[width=.95\linewidth]{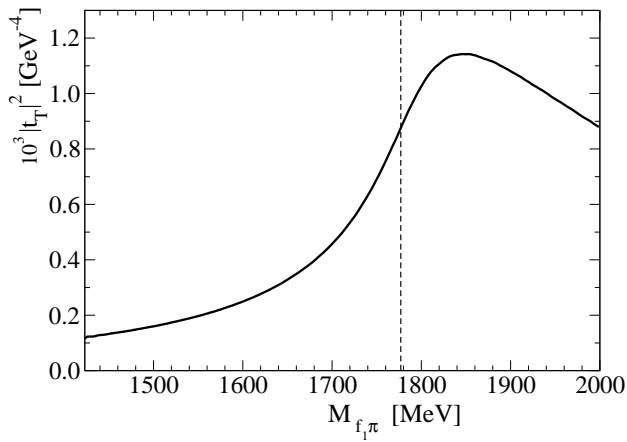}}\\
    \caption{Modulus squared of the triangle function, $|t_T|^2$,
   as a function of the $f_1\pi$ invariant mass. The vertical dashed line indicates the upper limit of the $f_1\pi$ distribution in the $\tau \to  f_1(1285)\pi\nu_\tau $ decay, which is $m_\tau=1777\mev$.}
    \label{fig:tT}
\end{figure}

\section{Results}

 \begin{figure}[tbp]
     \centering{
          \includegraphics[width=.98\linewidth]{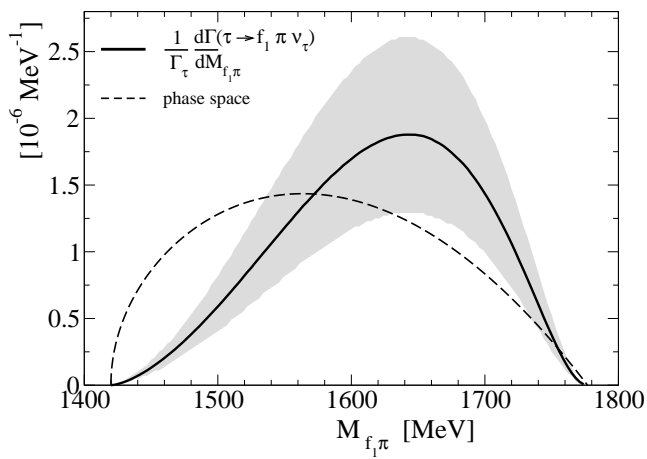}}\\
    \caption{Solid line: invariant mass distribution for the  $\tau \to  f_1(1285)\pi\nu_\tau $ decay divided by the total $\tau$ width. The dashed area represents the error band of the calculation. Dashed line: phase space normalized to match the area of the solid line. }
    \label{fig:InvMass}
\end{figure}

In Fig.~\ref{fig:tT} we plot the triangle $t_T$ as a function of the $f_1\pi$ invariant mass. In spite of the actual mass of the $f_1$ being about 100$\mev$ below the required mass to get a singularity and the broadening caused by the finite $K^*$ widths, we still see a prominent increase of the triangle function which then justifies the importance of the triangle mechanism in the present reaction, as anticipated in the discussion around  Eq.~\eqref{eq:qomenosqa}.
But in order to see its quantitative relevance we have to see its effect on the $\tau \to  f_1(1285)\pi\nu_\tau $ decay width. This is shown in  Fig.~\ref{fig:InvMass}, where we plot (solid line) the result  for the mass distribution of the branching ratio $\frac{1}{\Gamma_\tau}\frac{d\Gamma_{\tau\to f_1\pi \nu_\tau}}{dM_{f_1\pi}}$ including the error band calculated implementing a Monte Carlo gaussian sampling of the parameters within their uncertainties. The parameters with errors considered in the error analysis are: $R=3.0\pm 1.0\gev^{-1}$ (see Eq.~\eqref{eq:blatt}), $m_{K^ *}=894\pm2\mev$, $\Gamma_{K^*}=48\pm 2\gev$, $\textrm{BR}(\tau^-\to\nu_\tau {K^*}^-{K^*}^0)=(2.1\pm 0.5)\times 10^{-4}$ (see paragraph below Eq.~\eqref{eq:blatt}), $g=4.31\pm0.10$ (see Eq.~\eqref{eq:VPPlag}) and $g_{f_1}=7350\pm 130\mev$ (see discussion at the beginning of the Formalism section).
In Fig.~\ref{fig:InvMass} we also plot (dashed line) the phase space normalized to the area below the solid line. We can see that the strength of the distribution is clearly shifted towards the higher part of the spectrum, manifesting the important effect of the increasing tail of the triangle loop (see Fig.~\ref{fig:tT}).
This finding is interesting and the shifted peak is not tied to any resonance in the $f_1\pi$ system. In Ref.~\cite{Volkov:2018fyv} a possible shift of the strength of the invariant mass distribution to higher masses is tied to the  $a_1(1640)$ excitation. Even though their distribution peaks at lower energies than ours \cite{Volkov:2018fyv}.

The integrated area below the distribution provides the branching ratio for the 
$\tau\to f_1\pi \nu_\tau$ decay which gives
\be
BR(\tau\to f_1(1285) \pi \nu_\tau)= (3.6\pm 1.2)\times 10^{-4},
\ee
which compares well with the experimental value $(3.9\pm 0.5)\times 10^{-4}$ \cite{pdg} within uncertainties.
The agreement with the experimental result for the overall branching ratio is a significant and non-trivial result. First of all because the decay proceeds in our model through the triangle mechanism which has a non-trivial shape and strength. On the other hand, the absolute normalization of the branching ratio depends crucially on the coupling, $g_{f_1}$ of the  $\f$ to  $\bar K^*K$ (and $K^*\bar K$) which is a non-trivial prediction of the chiral unitary approach, and hence the result reinforces the idea of the dynamical, or molecular, nature of the $\f$ resonance. 
Experimental results on this distribution would be mostly welcome to support or disprove the previous claims.

\section*{Acknowledgments}

 This work is
partly supported by the Spanish Ministerio de Economia
y Competitividad and European FEDER funds under the
contract number FIS2014-57026-REDT, FIS2014-51948-
C2-1-P, FIS2014-51948-C2-2-P and FIS2017-84038-C2-2-P, and the Generalitat
Valenciana in the program Prometeo II-2014/068.

\appendix

\section{Quark spinor algebra}
\label{appendix:app}

In this Appendix we detail the evaluation of the  $W^-\to K^* \bar K^*$ vertex, the steps from Eq.~\eqref{eq:qqmat} to Eq.~\eqref{eq:qqmatred}, and the explicit steps that lead to Eq.~\eqref{eq:contr}.

We work in the rest of reference of the $W^-$ or total $K^* \bar K^*$ pair at rest. In that frame, and since in the present work we are going to be at an energy range where the invariant mass of the $K^* \bar K^*$ is of the order of $m_\tau \sim 2m_{K^*}$, the individual $K^*$ and $ \bar K^*$ are also having small three-momentum, $\vec q \sim 0$.
Thus 
\be
H^\mu=\bar u_d(\vec q)\,\gamma^\mu(1-\gamma_5)\,v_u(-\vec q) \simeq 
 \bar u_d(0)\,\gamma^\mu(1-\gamma_5)\,v_u(0),
\label{eq:qqmatapp}
\ee
with the Dirac 4-spinors
\be
u_d(0)=
\left(
\begin{array}{c}
\chi\\ 0 
\end{array}
\right) \qquad , 
v_u(0)=
\left(
\begin{array}{c}
0\\ \chi' 
\end{array}
\right).
\ee
In Eq.~\eqref{eq:qqmatapp} we need the following terms (where we use the Dirac representation of the $\gamma$ matrices)

\begin{align}
&\left(\chi,0\right )\gamma^0 \left(\begin{array}{c}0\\ \chi'\end{array}\right)=
\left(\chi,0\right )
\left(\begin{array}{c c}I&0\\ 0&-I\end{array}\right)
\left(\begin{array}{c}0\\ \chi'\end{array}\right)=
0 
\nn\\
&\left(\chi,0\right )\gamma^i \left(\begin{array}{c}0\\ \chi'\end{array}\right)=
\left(\chi,0\right )
\left(\begin{array}{c c}0&\sigma^i\\ -\sigma^i&0\end{array}\right)
\left(\begin{array}{c}0\\ \chi'\end{array}\right)=
(\chi,\sigma^i\chi')
\nn\\
&\left(\chi,0\right )\gamma^0\gamma^5 \left(\begin{array}{c}0\\ \chi'\end{array}\right)=
(\chi,0)
\left(\begin{array}{c c}I&0\\ 0&-I\end{array}\right)
\left(\begin{array}{c c}0&I\\ I& 0\end{array}\right)
\left(\begin{array}{c}0\\ \chi'\end{array}\right)=
(\chi,\chi')
\nn\\
&\left(\chi,0\right )\gamma^i\gamma^5 \left(\begin{array}{c}0\\ \chi'\end{array}\right)=
(\chi,0)
\left(\begin{array}{c c}0&\sigma^i\\ -\sigma^i&0\end{array}\right)
\left(\begin{array}{c c}0&I\\ I& 0\end{array}\right)
\left(\begin{array}{c}0\\ \chi'\end{array}\right)=
0
\label{eq:chis}
\end{align}
Therefore, only the $\gamma^i$ and $\gamma^0\gamma^5$ contribute.
As explained in the paragraph above Eq.~\eqref{eq:qqmat},
the initial $\bar u d$ pair must be in $L=1$, $S=1$, to match the parity and angular momentum of the final $K^* \bar K^*$ pair. 
The term with $\gamma^0\gamma^5$ in Eq.~\eqref{eq:chis} does not contribute to $S=1$. Indeed, $(\chi,\chi')=\delta_{s_1 s_2}$ projected over total spin $S=1$ gives
\be
\sum_{s_1}\delta_{s_1 s_2}{\cal C}(\frac{1}{2} , \frac{1}{2}, S;s_1, -s_2, 0)(-1)^{\frac{1}{2}+s_2}
\label{eq:sums1a}
\ee
where $s_1$ ($s_2$) is the spin third component of the $u$ ($\bar d$) quark and ${\cal C}(j_,j_2,J;m_1,m_2,M)$ the Clebsch-Gordan coefficient. Note the phase $(-1)^{\frac{1}{2}+s_2}$ coming from the hole interpretation of the antiparticle where $b^\dagger(\alpha,s,m_s)=(-1)^{s+m_s}a(\bar \alpha, s,-m_s)$ 
\cite{bohr}, where $b^\dagger$ is the antiparticle (hole) creation operator and $a$ the particle annihilator.
Thus, for $S=1$, Eq.~\eqref{eq:sums1a} is zero.
The only non-zero contribution for $S=1$ comes from the $(\chi,\sigma^i\chi')$
in Eq.~\eqref{eq:chis}. (We will use in the following $|s_1\rangle$ to stand for the spinor $\chi$ and $|s_2\rangle$ for $\chi'$).

The projection over spin $S$ gives, considering $\vec \sigma$ in the spherical basis and using the Wigner-Eckart theorem,
\begin{align}
&\sum_{s_1} \langle s_1| \sigma_\mu |s_2\rangle  {\cal C}(\frac{1}{2} , \frac{1}{2}, S;s_1, -s_2, s_1-s_2)(-1)^{\frac{1}{2}+s_2}
\nn \\
=&\sum_{s_1} {\cal C}(\frac{1}{2} ,1, \frac{1}{2};s_2, \mu, s_1)\langle\frac{1}{2}||\sigma|| \frac{1}{2}\rangle 
\nn \\ 
&\times {\cal C}(\frac{1}{2} , \frac{1}{2},S;s_1, -s_2, s_1-s_2)(-1)^{\frac{1}{2}+s_2}
\nn \\ 
=&\sum_{s_1} \sqrt{\frac{2}{3}} (-1)^{\frac{1}{2}-s_2} {\cal C}(\frac{1}{2} , \frac{1}{2}, 1;s_1, -s_2, \mu)\sqrt{3}
\nn \\ 
&\times
(-1)^{\frac{1}{2}+s_2} {\cal C}(\frac{1}{2} , \frac{1}{2},S;s_1, -s_2, s_1-s_2)
\nn \\ 
=&-\sqrt{2} \sum_{s_1} {\cal C}(\frac{1}{2} , \frac{1}{2}, 1;s_1, \mu-s_1, \mu){\cal C}(\frac{1}{2} , \frac{1}{2}, S;s_1, \mu-s_1, \mu)
\nn \\ 
=&-\sqrt{2}\, \delta_{S1} 
\label{eq:sums1b}
\end{align}
where we have used that $\langle\frac{1}{2}||\sigma|| \frac{1}{2}\rangle=\sqrt{3}$ and the property
\begin{align}
&{\cal C}(j_1,j_2,j_3;m_1,m_2,m_3) \nn\\
&=(-1)^{j_1-m_1}\left(\frac{2j_3+1}{2j_2+1}\right)^{\frac{1}{2}}{\cal C}(j_3,j_1,j_2;m_3,-m_1,m_2),
\label{eq:deltaS1}
\end{align}
and thus it gives contribution when $S=1$.
Note that the final result of Eq.~\eqref{eq:sums1b} does not depend on the third component, $\mu$, of the total spin, $S$.

For the evaluation of the amplitude squared we need the contraction with the leptonic part, (see Eq.~\eqref{eq:contr}), and, since we only have contribution from $\sigma_i$, we only need the space components of 
 Eq.~\eqref{eq:LL}.
We will first have terms that go as $\vec\sigma\cdot\vec p$ which contribute to $S=1$ as
\begin{align}
&\sum_{s_1} \langle s_1| \vec\sigma\cdot\vec p| s_2\rangle
 {\cal C}(\frac{1}{2} , \frac{1}{2},1;s_1, -s_2,s_1-s_2)(-1)^{\frac{1}{2}+s_2}
 \nn\\
 &=p\sum_{s_1} \langle s_1| \sigma_z| s_2\rangle
 {\cal C}(\frac{1}{2} , \frac{1}{2},1;s_1, -s_2,s_1-s_2)(-1)^{\frac{1}{2}+s_2}
 \nn\\
 &=-\sqrt{2}\,p 
\label{eq:sp}
\end{align}
where we have taken $\vec p$ in the $\hat z$ direction without loss of generality and put $\langle s_1| \sigma_z| s_2\rangle=(-1)^{\frac{1}{2}-s_2}\delta_{s_1 s_2}$.
We also have other contributions in Eq.~\eqref{eq:contr} that go as
 $p\cdot p'\vec\sigma\cdot\vec\sigma^*$ coming from the $- g^{\mu\nu}p'\cdot p$ part in Eq.~\eqref{eq:LL}, which will contribute to $S=1$ as
 \begin{align}
&\sum_{\mu}(-1)^\mu\sum_{s_1}\langle s_1|\sigma_\mu|s_2\rangle
{\cal C}(\frac{1}{2},\frac{1}{2},S;s_1, -s_2, s_1-s_2)(-1)^{\frac{1}{2}+s_2} 
 \nn\\
 &\times
 \sum_{s'_1}\langle s'_2|\sigma_{-\mu}|s'_1\rangle
{\cal C}(\frac{1}{2},\frac{1}{2},S;s'_1, -s'_2, s'_1-s'_2)(-1)^{\frac{1}{2}+s'_2}
 \nn\\
 &=\sum_{\mu}(-1)^\mu\sum_{s_1}
 \sqrt{3}  \,
 {\cal C}(\frac{1}{2}, 1,\frac{1}{2} ;s_2,\mu, s_1)
 \nn\\
&\times {\cal C}(\frac{1}{2},\frac{1}{2},S;s_1, -s_2, s_1-s_2)(-1)^{\frac{1}{2}+s_2} 
 \nn\\
&\times  \sum_{s'_1}\sqrt{3}  \,
 {\cal C}(\frac{1}{2} ,1, \frac{1}{2};s'_1, -\mu, s'_2)
 \nn\\
&\times 
 {\cal C}(\frac{1}{2},\frac{1}{2},S;s'_1, -s'_2, s'_1-s'_2)(-1)^{\frac{1}{2}+s'_2}
\nn\\
 &=
 \sum_{\mu}(-1)^\mu\sum_{s_1}\sqrt{3}\sqrt{\frac{2}{3}}\,
 {\cal C}(\frac{1}{2} , \frac{1}{2},1;s_1, -s_2, \mu)(-1)^{\frac{1}{2}-s_2}
 \nn\\
&\times {\cal C}(\frac{1}{2},\frac{1}{2},S;s_1, -s_2, s_1-s_2)(-1)^{\frac{1}{2}+s_2} 
 \nn\\
&\times  \sum_{s'_1} \sqrt{3} \sqrt{\frac{2}{3}} \,
 {\cal C}(\frac{1}{2} ,\frac{1}{2},1;s'_1, -s'_2,\mu)(-1)^{\frac{1}{2}-s'_1}
 \nn\\
 &\times{\cal C}(\frac{1}{2},\frac{1}{2},S;s'_1, -s'_2, s'_1-s'_2)(-1)^{\frac{1}{2}+s'_2}
 \nn\\
 &=
2 \sum_{\mu}  \sum_{s_1}{\cal C}(\frac{1}{2} , \frac{1}{2},1;s_1, -s_2, \mu)
{\cal C}(\frac{1}{2},\frac{1}{2},S;s_1, -s_2, s_1-s_2)
\nn\\
&\times\sum_{s'_1}
{\cal C}(\frac{1}{2} ,\frac{1}{2},1;s'_1, -s'_2,\mu)
{\cal C}(\frac{1}{2},\frac{1}{2},S;s'_1, -s'_2, s'_1-s'_2)
\nn\\
 &=2 \sum_{\mu=-1}^{+1} \delta_{S1}\delta_{S1}=6 \,\delta_{S1}.
 \label{eq:tocho}
 \end{align}

Finally, the term $\epsilon^{\alpha\mu\beta\nu}p'_\alpha p_\beta$ in 
Eq.~\eqref{eq:LL} gives zero contribution. Indeed, only space components for $\mu$ and $\nu$ are possible since it is contracted with $\sigma^i\sigma^j$ for Eq.~\eqref{eq:contr}. In addition, since $\vec p=\vec p\,'$ in the reference frame we are working on, if  $\alpha$ and  $\beta$ are both spatial then it would be zero since we would be contracting an antisymmetric tensor with a symmetric one. Therefore the only possible non-zero contribution would come from $\epsilon^{0ikj}p'_0 p_k$ or $\epsilon^{ki0j}p'_k p_0$. Since $\epsilon^{0ijk}=\epsilon_{ijk}$ we have to evaluate something proportional to 
\begin{align}
&\epsilon_{ijk} \sigma_i p_j \sigma_k^*= \vec p \cdot (\vec\sigma\times\vec\sigma^*)
\nn\\
&= -\sqrt{2}i\sum_{\nu}(-1)^{\nu+\mu}\sum_\mu{\cal C}(1,1,1,;\mu,\nu,\mu+\nu)\sigma_\mu\sigma_\nu^*\,p_{-\nu-\mu}
\label{eq:ssp}
\end{align}
where we have used the spherical basis for the vectors.
Considering  $\vec p$ in the $\hat z$ direction and, then, $\mu+\nu=0$, Eq.~\eqref{eq:ssp} reads
\begin{align}
 -\sqrt{2}i\,p
 \sum_\mu{\cal C}(1,1,1,;\mu,-\mu,0)\sigma_\mu\sigma_{-\mu}^*
\label{eq:ssp2}.
\end{align}
When applying the matrix element $\langle s_1|\cdots |s_2\rangle$ projected over spin $S=1$ to Eq.~\eqref{eq:ssp2}, 
and using $\langle s'_2|\sigma_{-\mu}|s'_1\rangle
=(-1)^{\mu}\langle s'_1|\sigma_{\mu}|s'_2\rangle$,
 we would have, up to the global factor $-\sqrt{2}ip$, the same as in 
Eq.~\eqref{eq:tocho} but having ${\cal C}(1,1,1,;\mu,-\mu,0)$ instead of $(-1)^\mu$. Therefore we would arrive to the same last line of Eq.~\eqref{eq:tocho} except that now it would be
\begin{align}
&-2\sqrt{2}ip  \sum_{\mu=-1}^{+1} (-1)^\mu{\cal C}(1,1,1,;\mu,-\mu,0)\delta_{S1}
\delta_{S1}
\nn\\
&=2\sqrt{2}ip\, \delta_{S1}\left(\frac{1}{\sqrt{2}}-\frac{1}{\sqrt{2}}\right)
=0.
\label{eq:ssp3}
\end{align}
And hence the term  $\epsilon^{\alpha\mu\beta\nu}p'_\alpha p_\beta$ in 
Eq.~\eqref{eq:LL} does not contribute.

Taking into account the results from Eqs.~\eqref{eq:sp}, \eqref{eq:tocho} and
\eqref{eq:ssp3}, the contraction between the leptonic and hadronic part of Eq.~\eqref{eq:contr} is
\begin{align}
&
\overline \sum L^\mu {L^\nu}^\dagger H_\mu H_\nu^\dagger
\nn\\
&=
4[ (-\sqrt{2})|\vec p\,'|(-\sqrt{2})|\vec p|
+  (-\sqrt{2})|\vec p |(-\sqrt{2})|\vec p\,'|
+6 p\cdot p']\nn\\
&=
24\left( E_\tau E_\nu -\frac{\vec p\,^2}{3}\right).
\label{eq:contrapp}
\end{align}

\end{document}